\documentclass[twocolumn,prl,aps,amsfonts,showpacs]{revtex4}

\usepackage{graphicx}

\begin{document}
\preprint{}

\newcommand{\1}{{\bf \scriptstyle 1}\!\!{1}}
\newcommand{\I}{{\rm i}}
\newcommand{\p}{\partial}
\newcommand{\D}{^{\dagger}}
\newcommand{\bx}{{\bf x}}
\newcommand{\br}{{\bf r}}
\newcommand{\bk}{{\bf k}}
\newcommand{\bv}{{\bf v}}
\newcommand{\bp}{{\bf p}}
\newcommand{\bs}{{\bf s}}
\newcommand{\bu}{{\bf u}}
\newcommand{\bA}{{\bf A}}
\newcommand{\bB}{{\bf B}}
\newcommand{\bE}{{\bf E}}
\newcommand{\bF}{{\bf F}}
\newcommand{\bI}{{\bf I}}
\newcommand{\bK}{{\bf K}}
\newcommand{\bL}{{\bf L}}
\newcommand{\bP}{{\bf P}}
\newcommand{\bQ}{{\bf Q}}
\newcommand{\bS}{{\bf S}}
\newcommand{\bH}{{\bf H}}
\newcommand{\balpha}{\mbox{\boldmath $\alpha$}}
\newcommand{\bsigma}{\mbox{\boldmath $\sigma$}}
\newcommand{\bSigma}{\mbox{\boldmath $\Sigma$}}
\newcommand{\bOmega}{\mbox{\boldmath $\Omega$}}
\newcommand{\bpi}{\mbox{\boldmath $\pi$}}
\newcommand{\bphi}{\mbox{\boldmath $\phi$}}
\newcommand{\bnabla}{\mbox{\boldmath $\nabla$}}
\newcommand{\bmu}{\mbox{\boldmath $\mu$}}
\newcommand{\bepsilon}{\mbox{\boldmath $\epsilon$}}

\newcommand{\iLambda}{{\it \Lambda}}
\newcommand{\cA}{{\cal A}}
\newcommand{\cD}{{\cal D}}
\newcommand{\cF}{{\cal F}}
\newcommand{\cL}{{\cal L}}
\newcommand{\cH}{{\cal H}}
\newcommand{\cI}{{\cal I}}
\newcommand{\cM}{{\cal M}}
\newcommand{\cO}{{\cal O}}
\newcommand{\cR}{{\cal R}}
\newcommand{\cU}{{\cal U}}
\newcommand{\cT}{{\cal T}}

\newcommand{\be}{\begin{equation}}
\newcommand{\ee}{\end{equation}}
\newcommand{\bea}{\begin{eqnarray}}
\newcommand{\eea}{\end{eqnarray}}
\newcommand{\beqa}{\begin{eqnarray*}}
\newcommand{\eeqa}{\end{eqnarray*}}
\newcommand{\nn}{\nonumber}
\newcommand{\DD}{\displaystyle}

\newcommand{\ba}{\left[\begin{array}{c}}
\newcommand{\baa}{\left[\begin{array}{cc}}
\newcommand{\baaa}{\left[\begin{array}{ccc}}
\newcommand{\baaaa}{\left[\begin{array}{cccc}}
\newcommand{\ea}{\end{array}\right]}


\title{Berry-phase blockade in single-molecule magnets}

\author{Gabriel Gonz\'{a}lez$^{1,2}$ and Michael N.~Leuenberger$^{1,2}$}\email{mleuenbe@mail.ucf.edu}

\affiliation{$^1$NanoScience Technology Center, University of Central
Florida, Orlando, FL 32826} 

\affiliation{$^2$Department of Physics, University of Central
Florida, Orlando, FL 32816}

\date{\today}

\begin{abstract}
We formulate the problem of electron transport through a single-molecule magnet (SMM) in
the Coulomb blockade regime taking into account topological interference effects for the tunneling of the large spin of a SMM. The interference originates from spin Berry phases associated
with different tunneling paths. We show that in the case of incoherent spin states it is essential to place the SMM between oppositely spin-polarized source and drain leads in order to detect the spin tunneling in the stationary current, which exhibits topological zeros as a function of the transverse magnetic field.
\end{abstract}


\maketitle

Single-molecule magnets (SMMs), such as Mn$_{12}$ \cite{experiments,delBarco} and Fe$_8$
\cite{Sangregorio,Wernsdorfer}, have become the focus of intense research in the last decade since experiments on bulk samples demonstrated
the quantum tunneling of a single magnetic moment on a macroscopic scale.
These molecules are characterized by a large total spin, a large magnetic
anisotropy barrier, and anisotropy terms which allow the spin to tunnel through the barrier. Transport through SMM offers several unique features with potentially large impact in applications for magnetic devices based on SMM such as high-density magnetic storage as well as
quantum computing applications \cite{app}. Recently experiments have pointed out the importance of the interference between spin tunneling paths in molecules and its effects on electron transport scenarios involving SMMs. For instance,
measurements of the magnetization in bulk F$_8$ samples (see
Ref.~\cite{Zener}) have observed oscillations in the tunnel splitting
$\Delta E_{s,-s}$ between states $S_z = s$ and $-s$ as a function of a
transverse magnetic field at temperatures between $0.05$ K and $0.7$
K. This effect can be explained by the interference between Berry phases acquired by spin tunneling paths of opposite windings using a coherent spin-state path integral approach \cite{Loss,Delft,Garg,Leuenberger_Berry}. \\
The ability to manufacture molecular structures with fixed magnetic properties has given rise to a field known as molecular magnetism. To date, several experiments have shown the possibility to work with an individual SMM preserving the magnetic properties \cite{newexps}, thereby demonstrating the Coulomb blockade effect at low temperature in a single SMM transistor geometry \cite{newexps}. The theoretically predicted Kondo effect in SMMs \cite{Romeike_Kondo,future} has not been observed yet.
A theoretical description of the observed Coulomb blockade effect has recently been given by means of {\it coherent} spin states in Ref. \cite{Romeike_blockade}. \\
In this letter we propose the Berry-phase blockade effect by coupling an individual SMM
to spin-polarized leads. We analyze the transport properties of the system in the 
Coulomb blockade regime for the ground state of a SMM in the presence of a longitudinal and transverse magnetic field. 
Since the decoherence time between degenerate spin states can be as low as $T_2=10^{-9}\,s$ \cite{delBarco2005,Hill}, we work with {\it incoherent} spin states. We  show that in the case of {\it incoherent} spin tunneling it is essential to use oppositely spin-polarized source and drain leads in order to be able to observe variations of the stationary current as a function of longitudinal or transverse magnetic field. In particular, the current can be suppressed due to the Berry-phase interference of the spin tunneling paths. In the case of fully polarized leads, complete current suppression coincides precisely with the topological zeros of the spin tunneling. Even in the case of partially polarized leads there are still fingerprints of the Berry-phase blockade which lead to significant changes in the stationary current through the SMM.   \\
In the following we present our calculations that are valid for any SMM in the Coulomb blockade regime coupled to polarized leads and in the presence of a longitudinal and transverse magnetic field. We derive the (generalized) master equation for the low energy states and calculate the stationary current through the SMM for the cases of unpolarized, fully, and partially polarized leads. We apply our results to the newly synthesized SMM Ni$_{4}$, which has a spin of $s=4$ and a ground state tunnel splitting of $\Delta E_{s,-s}\approx 0.01$\ K at zero magnetic field \cite{Cber}, corresponding to an angular frequency of $\omega=10^{9}$ s$^{-1}$.\\
Consider a SMM in the Coulomb blockade regime which is tunnel-coupled to two polarized leads at the chemical potentials $\mu_{l}$, where $l=L,R$ denote the left and right lead, respectively. The total Hamiltonian is given by
\be
H=H_{l}+\cH_{\rm s}+H_{m},
\label{H}
\ee
where $H_{l}=\sum_{lk\sigma}\epsilon_{lk}c^{\dagger}_{lk\sigma}c_{lk\sigma}$ represents the energy of the leads. $c^{\dagger}_{lk\sigma}$ creates an electron in lead $l$ with orbital state $k$, spin $\sigma$,and energy $\epsilon_{lk}$. The coupling of the leads to the molecule is described by $H_{m}=\sum_{lpk\sigma}t^{\sigma}_{lp}c^{\dagger}_{lk\sigma}d_{p\sigma}+H.c.$, where $t^{\sigma}_{lp}$ denotes the tunneling amplitude and $d^{\dagger}_{p\sigma}$ creates an electron on the molecule in orbital state $p$. The term $\cH_{\rm spin}$ is typically given by the spin Hamiltonian of a SMM in an external transverse magnetic field $H_\bot$ and a longitudinal magnetic field $H_{z}$, i.e. 
\bea
\cH_{\rm s} &=& -A_qS_{q,z}^2+\frac{B_q}{2} \left( S_{q,+}^2
+ S_{q,-}^2 \right)+\frac{B_{4,q}}{3} \left( S_{q,+}^4 + S_{q,-}^4\right) \nn\\
& &  +g\mu_B H_{z}S_{q,z}+\frac{1}{2}(h_\bot^* S_{q,+}+h_\bot S_{q,-}),
\eea
where the easy axis is taken along $z$, $S_{q,\pm} = S_{q,x} \pm
iS_{q,y}$, and the integer index $q$ denotes the charging state of the
SMM, where $q=-1$ adds one electron to the molecule and $q=0$ when the molecule is 
neutral. The transverse magnetic field $H_\bot = H_x + iH_y = |H_\bot|
e^{i\varphi}$ lies in the $xy$ plane. We use the abbreviation
$h_\bot=g\mu_B H_\bot$. In this Hamiltonian, the dominant longitudinal
anisotropy term creates a ladder structure in the molecule spectrum
where the eigenstates $|\pm m_q\rangle$ of $S_z$ are degenerate. The
weaker transverse anisotropy terms couple these states. The coupling
parameters depend on the charging state of the molecule. For example,
it is known that Mn$_{12}$ changes its easy-axis anisotropy constant
(and its total spin) from $A_0=56$ $\mu$eV ($S_0=10$) to $A_{-1}=43$
$\mu$eV ($S_{-1}=19/2$) and $A_{-2}=32$ $\mu$eV ($S_{-2}=10$) when
singly and doubly charged, respectively \cite{dataexps}.
Experiments with Ni$_4$ show that $B_{4,q=0}=-0.003$ K,
i.e. $B_{4,0}$ is negative \cite{Cber}. In this case, in order to see the Berry
phase oscillation, a magnetic field $H_\bot$ must be applied in the
$xy$ plane \cite{Leuenberger_Berry} along specific angles
$\varphi(B_q,B_{4,q})$. It is also possible to tune the tunnel splitting by means of the longitudinal magnetic field $H_{z}$. \\
For weak coupling between the leads and the SMM we use the standard formalism suitable to describe a system (SMM) coupled to a reservoir (polarized leads)\cite{Blum}. The master equation describing the electronic spin states of the SMM is given in Born and Markov approximation by
\be
\dot{\rho}_{m,n}=\frac{i}{\hbar}[\rho,H]_{m,n}+\delta_{m,n}\sum_{l\neq m}\rho_{n} W_{m,l}-\gamma_{m,n}\rho_{m,n},
\label{Mastereq}
\ee
where $\gamma_{m,n}=\frac{1}{2}\sum_{l}(W_{l, n}+W_{l, m})+1/T_{2}$ is the total decoherence rate which contains the spin decoherence time $T_{2}$ due to e.g. nuclear spins and the rates $W_{m, n}$ of transition between the states of the SMM. Figs. (\ref{Diagram2}) and (\ref{Diagram3}) show the ground states  $s$ and $-s$ of a SMM placed between unpolarized and polarized leads, respectively. $w_{\downarrow\uparrow}^{(l)}$ represents the spin-dependent transition rate from the $l=L,R$ lead to the SMM and are defined in Fermi's golden rule approximation by $w_{\downarrow}^{(l)}=2\pi{\it D}\nu^{(l)}_{\downarrow}|t^{(l)}_{\downarrow}|^2 /\hbar$ and 
$w_{\uparrow}^{(l)}=2\pi{\it D}\nu^{(l)}_{\uparrow}|t^{(l)}_{\uparrow}|^2 /\hbar$, respectively, where ${\it D}$ is the density of states and $\nu^{(l)}_{\uparrow}$ and 
$\nu^{(l)}_{\downarrow}$ are fractions of the number of spins polarized up and down of lead $l$ such that $\nu^{(l)}_{\downarrow}+\nu^{(l)}_{\uparrow}=1$. $t^{(l)}_{\uparrow}$ and $t^{(l)}_{\downarrow}$ are  the tunneling amplitudes of lead $l$, respectively. Typical values for the tunneling rate of the electron range from around $w=10^6$ s$^{-1}$ to $w=10^{10}$ s$^{-1}$ (see Refs.~\cite{newexps,park}).

In order to see coherent spin tunneling, the decoherence time must be increased for example by deuteration\cite{Wernsdorfer2000} such that $1/T_2\ll \Delta E_{s,-s}/\hbar$ and at the same time the contact to the leads must be so weak that $W_{m, n}\ll \Delta E_{s,-s}/\hbar$. Another possibility is to increase the transverse magnetic field $|H_\bot|$ beyond the Berry-phase oscillations. In this case unpolarized leads can be used to measure the tunnel splitting between the coherent spin states $(\left|s'\right\rangle+\left|-s'\right\rangle)/\sqrt{2}$ and $(\left|s'\right\rangle-\left|-s'\right\rangle)/\sqrt{2}$ by varying the gate or bias voltage. However, only partially or fully polarized leads allow us to probe the incoherent tunneling rate $\Gamma_{s,-s}$ between the ground states $s$ and $-s$ for $q=0$ and also between $s'$ and $-s'$ for $q=-1$. As we prove below, both incoherent tunneling rates $\Gamma_{s,-s}$ and $\Gamma_{s',-s'}$ contribute to the total polarized current through the SMM.
The sequential tunneling rates for absorption of an electron in Eq. (\ref{Mastereq}) for ground states with spin $s$ and $s'$ and energy differences $\Delta_{\pm s', \pm s}$ in the case of low temperatures are given by
\begin{equation}
\begin{array}{l}
W_{s', s}  =  \sum_{l}W^{(l)}_{s', s}\,, \qquad W^{(l)}_{s', s} = w_{\downarrow}^{(l)} f_{l}(\Delta_{s', s})\,,\\
W_{-s', -s}=  \sum_{l}W^{(l)}_{-s', -s} \,, W^{(l)}_{-s', -s}  =  w_{\uparrow}^{(l)} f_{l}(\Delta_{-s', -s})\,,
\end{array}
\end{equation}
and the tunneling rates for the emission of an electron are given by
\begin{equation}
\begin{array}{l}
W_{s,s'}    =  \sum_{l}W^{(l)}_{s,s'} \,,\qquad W^{(l)}_{s,s'}      =   w_{\downarrow}^{(l)} [1-f_{l}(\Delta_{s,s'})]\,,\\
W_{-s,-s'}  =  \sum_{l}W^{(l)}_{-s,-s'} \,, W^{(l)}_{-s,-s'}  =  w_{\uparrow}^{(l)} [1-f_{l}(\Delta_{-s,-s'})]\,,
\end{array}
\end{equation}
where $f_{l}(\Delta_{s', s})=[1+e^{(\Delta_{s', s}-\mu_{l})/kT}]^{-1}$ is the Fermi function.
The diagonal elements of (\ref{Mastereq}) yield
\be
\dot{\rho}_{s}=\frac{i}{\hbar}[\rho,H]_{s, s}+\sum_{n\neq s}\rho_{n} W_{s,n}-\rho_{s}\sum_{n\neq s}W_{n,s},
\label{delements}
\ee
and the off-diagonal elements of (\ref{Mastereq}) yield
\be
\dot{\rho}_{s,s'}=\frac{i}{\hbar}[\rho,H]_{s ,s'}-\gamma_{s,s'}\rho_{s,s'}.
\label{oelements}
\ee
Since we are interested in the long time behavior $t\gg 1/\gamma_{m,n}$, we can set $\dot{\rho}_{s,s'}=0$ in equation (\ref{oelements}) to obtain the following coupled differential equations for the diagonal elements of the density matrix
\begin{equation}
\begin{array}{l}
\dot{\rho}_{s} =  \left(\frac{\Delta E_{s, -s}}{2\hbar}\right)^2\frac{2\gamma_{s,-s}} {(H_{s}-H_{-s})^2/\hbar^2+\gamma_{s, -s}^2}(\rho_{-s}-\rho_{s}) \\
\qquad+W_{s,s'}\rho_{s'}-W_{s', s}\rho_{s},
\end{array}
\end{equation}
\begin{equation}
\begin{array}{l}
\dot{\rho}_{-s}  =  \left(\frac{\Delta E_{s, -s}}{2\hbar}\right)^2\frac{2\gamma_{s, -s}}{(H_{-s}-H_{s})^2/\hbar^2 +\gamma_{s, -s}^2}(\rho_{s}-\rho_{-s})\\
\qquad+W_{-s, -s'}\rho_{-s'}-W_{-s', -s}\rho_{-s}.
\end{array}
\end{equation}
The other two differential equations are obtained by just replacing $s \leftrightarrow s'$ in the above equations. Solving the set of differential equations  for $\rho_{s}$, $\rho_{-s}$,
$\rho_{s'}$ and $\rho_{-s'}$ in the stationary case ($t\gg 1/W_{m,n}$) we obtain 
\begin{equation}
\begin{array}{l}
\rho_{s}=(W_{s,s'}(W_{-s', -s}+\Gamma_{s, -s})\Gamma_{s', -s'}\\
					\qquad +W_{-s, -s'}(W_{s', s}+\Gamma_{s', -s'})\Gamma_{ s, -s})/\eta,\\
\rho_{-s}=(W_{s,s'}(W_{-s, -s'}+\Gamma_{s', -s'})\Gamma_{s, -s}\\
\qquad +W_{-s, -s'}(W_{s', s}+\Gamma_{s, -s})\Gamma_{s', -s'})/\eta,\\
\rho_{s'}=(W_{s', s}(W_{- s ,-s'}+\Gamma_{s', -s'})\Gamma_{s, -s}\\
\qquad +W_{-s', -s}(W_{s, s'}+\Gamma_{s, -s})\Gamma_{s', -s'})/\eta,\\
\rho_{-s'}=(W_{s', s}(W_{-s', -s}+\Gamma_{s ,-s})\Gamma_{s', -s'}\\
\qquad +W_{-s', -s}(W_{s, s'}+\Gamma_{s', -s'})\Gamma_{s, -s})/\eta,
\end{array}
\end{equation}
where $\eta$ is a normalization factor given by 
\be
\begin{array}{l}
\eta =(\Gamma_{4,-4}+\Gamma_{7/2,-7/2})(W_{-4,-7/2}W_{7/2,4}\\
+W_{4,7/2}W_{-7/2,-4})+2[\Gamma_{4,-4}W_{-4,-7/2}W_{4,7/2}\\
+\Gamma_{7/2,-7/2}W_{-7/2,-4}W_{7/2,4}+\Gamma_{4,-4}\Gamma_{7/2,-7/2}\\
(W_{7/2,4}+W_{-7/2,-4}+W_{4,7/2}+W_{-4,-7/2})],
\label{normalization}
\end{array}
\ee
such that $\sum_{n}\rho_{n}=1$. The incoherent tunneling rate is
\be
\Gamma_{s, -s}=\left(\frac{\Delta E_{s, -s}}{2\hbar}\right)^{2}\frac{2\gamma_{s, -s}}{\left( H_{s}-H_{-s}\right)^2/\hbar^2+\gamma_{s, -s}^2},
\label{gamma}
\ee
which is a Lorentzian as a function of the longitudinal Zeeman splitting $H_{s}-H_{-s}=g\mu_B H_{z}[s-(-s)]$. 
We now proceed to define the current through the SMM in terms of the density matrix elements. In the case of Ni$_4$ we have $s=4$ and $s'=7/2$, therefore the current reads
\be
I=e\left(W_{4, 7/2}\rho_{7/2}+W_{-4, -7/2}\rho_{-7/2}\right).
\label{spincurrent}
\ee
Taking into consideration the asymmetry of the leads ,i.e. $w_{\downarrow\uparrow}^{(L)}\neq w_{\downarrow\uparrow}^{(R)}$, and restricting ourselves to the case of unpolarized leads, i.e. $\nu^{(L)}_{\uparrow}=\nu^{(L)}_{\downarrow}=\nu^{(R)}_{\downarrow}=\nu^{(R)}_{\uparrow}=1/2$, we obtain the following conditions for the transition rates 
\be
W_{7/2,4}=W_{-7/2,-4},\qquad W_{4,7/2}=W_{-4,-7/2}.
\label{restriction1}
\ee
Substituting the values of $\rho_{7/2}$ and $\rho_{-7/2}$ into Eq. (\ref{spincurrent}) and using Eq. (\ref{restriction1}) we obtain 
\be
\frac{e}{I_{unp}}=\frac{1}{W_{7/2,4}}+\frac{1}{W_{-4,-7/2}},
\label{tranrates}
\ee
which does not depend on the tunnel splitting energy of the SMM. Thus it is impossible to observe Berry-phase oscillations for the case of unpolarized leads and incoherent spin states. Eq. (\ref{tranrates}) can be interpreted as two resistances in series \cite{ML}, where the only transitions that contribute to the current through the SMM are $4\leftrightarrow 7/2$ and $-4\leftrightarrow -7/2$ (see Fig.~\ref{Diagram2}).
\begin{figure}[htb]
\includegraphics[width=7.5cm]{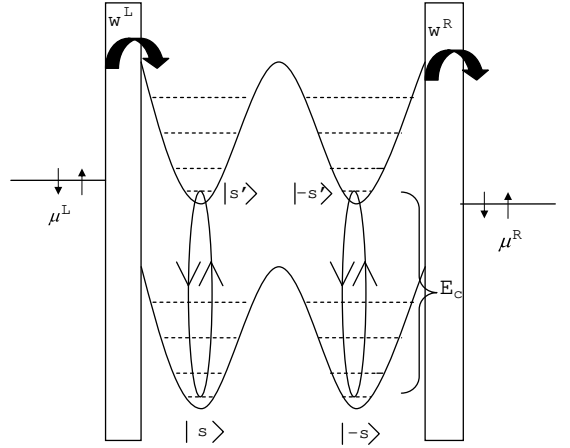}
\caption[]{Diagram for the transitions between the ground states $4\leftrightarrow 7/2$ and $-4\leftrightarrow -7/2$ in the case of unpolarized leads. The transitions arise from the sequential tunneling of unpolarized electrons in and out of the SMM. E$_{c}$ is the charging energy.}
\label{Diagram2}
\end{figure}
\\
In the case of leads that are fully polarized in opposite directions, i.e. $\nu^{(L)}_{\uparrow}=\nu^{(R)}_{\downarrow}=1$ or $\nu^{(L)}_{\downarrow}=\nu^{(R)}_{\uparrow}=1$, we get the following conditions for the transition rates:
\be
W_{-4,-7/2}=W_{7/2,4}=0,
W_{4,7/2}=W_{-7/2,-4}=0. 
\label{restriction2}
\ee
Choosing the case $\nu^{(L)}_{\uparrow}=\nu^{(R)}_{\downarrow}=1$ and using Eq. (\ref{restriction2}) we can then substitute the values of $\rho_{7/2}$ and $\rho_{-7/2}$ into Eq. (\ref{spincurrent}) to obtain
\be
\frac{e}{I_{p}}=\frac{2}{W_{-7/2, -4}}+\frac{1}{\Gamma_{4, -4}}+\frac{2}{W_{4, 7/2}}+\frac{1}{\Gamma_{7/2, -7/2}},
\label{spincurrent1}
\ee
which reflects the fact that the current through the SMM depends on the tunnel splittings and can be interpreted as four resistances coupled in series in a loop (see Fig.~\ref{Diagram3}).
\begin{figure}[htb]
\includegraphics[width=7.5cm]{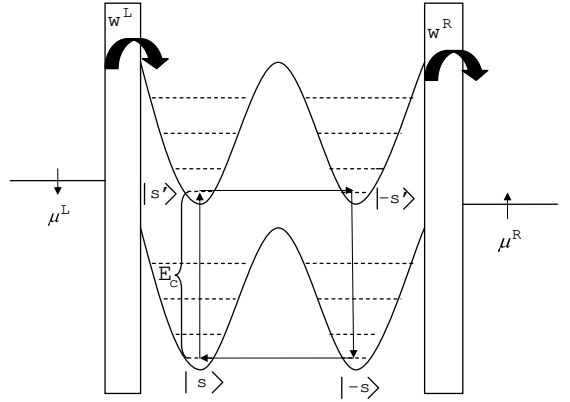}
\caption[]{Diagram for the transitions between the ground states $4\rightarrow 7/2\rightarrow -7/2 \rightarrow -4$ in the case of fully polarized leads in opposite directions $\nu^{(L)}_{\uparrow}=\nu^{(R)}_{\downarrow}=1$.}
\label{Diagram3}
\end{figure}
Notice the clockwise direction of the transition rates between the different states $s$, $s'$,$-s'$ and $-s$ of the SMM. If we chose to work with fully polarized leads of the form $\nu^{(L)}_{\downarrow}=\nu^{(R)}_{\uparrow}=1$, then the direction of the transition rates between states would be opposite, i.e. anticlockwise. \\
Fig.~(\ref{Polarizedleads1}) shows the current as a function of the transverse magnetic field $H_\bot$ for fully polarized leads. If the tunnel splitting $\Delta E_{4, -4}$ or $\Delta E_{7/2, -7/2}$ is topologically quenched, then $\Gamma_{4, -4}$ or $\Gamma_{7/2, -7/2}$ vanishes [see Eq. (\ref{gamma})], which leads to complete current suppression according to Eq. (\ref{spincurrent1}). Since this current blockade is a consequence of the topologically quenched tunnel splitting, we call it Berry-phase blockade. Note that the current can also be suppressed by applying the longitudinal magnetic field $H_{z}$ which follow immediately from Eqs. (\ref{gamma}) and (\ref{spincurrent1}).
\begin{figure}[htb]
\includegraphics[width=7.5cm]{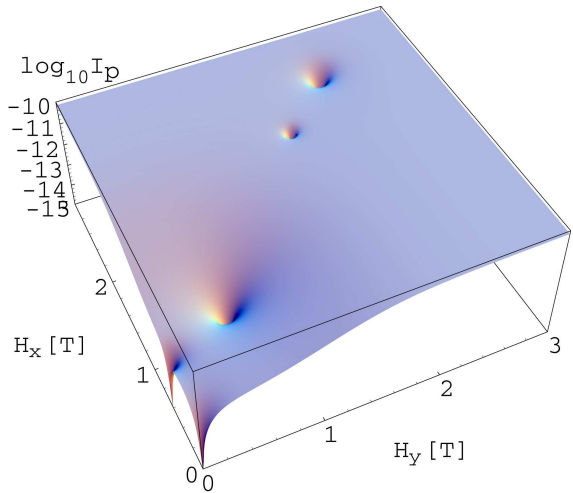}
\caption[]{The graph shows the logarithm base 10 of the stationary current versus the transverse magnetic field $H_\bot$ for $\nu^{(R)}_{\uparrow}=\nu^{(L)}_{\downarrow}=1$, $\nu^{(R)}_{\downarrow}=\nu^{(L)}_{\uparrow}=0$, $w_{\downarrow}^{(L)}=1\times 10^{9} s^{-1}$ and $w_{\downarrow}^{(R)}=10 w_{\downarrow}^{(L)}$.The polarizations of the left and right lead are given by $P^{R}=-P^{L}=\nu^{R}_{\uparrow}-\nu^{R}_{\downarrow}=100\, \%$. At the zeros of the tunnel splitting $\Delta E_{s, -s}$ or $\Delta E_{s', -s'}$ the current is completely suppressed. The scale varies from I$_{p}=0.1$\ nA to 1~fA.}
\label{Polarizedleads1}
\end{figure}
\\
If we consider now partially polarized leads (i.e. $\nu^{(R)}_{\uparrow}>\nu^{(R)}_{\downarrow} ,\, \nu^{(L)}_{\downarrow}>\nu^{(L)}_{\uparrow}$) and calculate the current through the SMM, we obtain
\begin{equation}
\begin{array}{l}
I_{pp}=e[\Gamma_{4,-4}W_{-4,-7/2}W_{4,7/2}(W_{-7/4,-4}+W_{7/2,4})\\
+\Gamma_{4,-4}\Gamma_{7/2,-7/2}(W_{-4,-7/2}+W_{4,7/2})(W_{-7/4,-4}+W_{7/2,4})\\
+\Gamma_{7/2,-7/2}W_{-7/4,-4}W_{7/2,4}(W_{-4,-7/2}+W_{4,7/2})]/\eta.
\end{array}
\end{equation}
Fig.~\ref{Polarizedleads2} shows the current as a function of the transverse magnetic field $H_\bot$ for partially polarized leads. The Berry-phase blockade is still visible in the stationary current even at a spin polarization of 60 $\%$, which makes it experimentally accessible since recent experiments have achieved near 100 $\%$ spin polarization \cite{Xiong}.\\ 
\begin{figure}[htb]
\includegraphics[width=7.5cm]{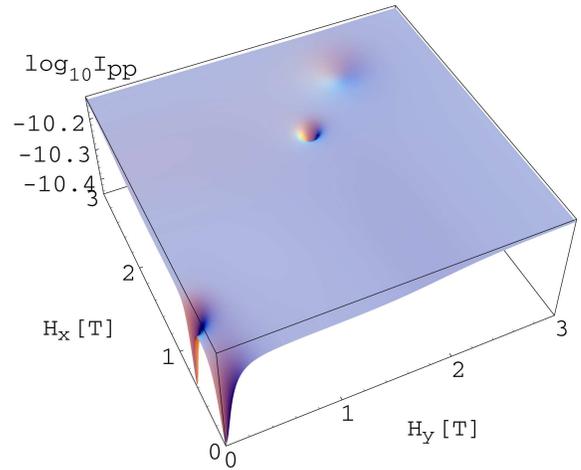}
\caption[]{The graph shows the logarithm of the stationary current versus the transverse magnetic field $H_\bot$ for $\nu^{(R)}_{\uparrow}=\nu^{(L)}_{\downarrow}=0.8$, $\nu^{(R)}_{\downarrow}=\nu^{(L)}_{\uparrow}=0.2$, $w_{\downarrow}^{(L)}=1\times 10^{9} s^{-1}$, and $w_{\downarrow}^{(R)}=10 w_{\downarrow}^{(L)}$. The polarizations of the left and right lead are given by $P^{R}=-P^{L}=\nu^{R}_{\uparrow}-\nu^{R}_{\downarrow}=60\, \%$. In this case at the zeros of the tunnel splitting $\Delta E_{s, -s}$ and $\Delta E_{s', -s'}$ the current is suppressed approximately by a factor of 3.}
\label{Polarizedleads2}
\end{figure}
\\
In summary, we have shown the Berry-phase blockade for a SMM placed between polarized leads. This behavior is due to Berry-phase interference of the SMM spin between different tunneling paths. We have shown that in the case of incoherent spin states it is essential to use polarized leads in order to observe the Berry-phase blockade. 

{\it Acknowledgment}. We thank E. del Barco, S. Khondaker and E. Mucciolo for useful discussions.



\end{document}